\documentclass{appolb}

\usepackage{epsfig}
\usepackage{graphicx}
\begin{document}

\title{The QCD Critical End Point in the Context of the
Polyakov--Nambu--Jona-Lasinio Model
\thanks{Presented at SQM2011}%
}

\author{P. Costa, C. A. de Sousa, M. C. Ruivo
\address{Centro de F\'isica Computacional, Departamento de F\'isica,
Universidade de Coimbra, P-3004-516 Coimbra, Portugal}
\and
{H. Hansen}
\address{IPNL, Universit\'e de Lyon/Universit\'e Lyon 1, CNRS/IN2P3, 4 rue E.Fermi, F-69622
Villeurbanne Cedex, France}
}
\maketitle


\begin{abstract}

We investigate the phase diagram of the so-called Polyakov--Nambu--Jona-Lasinio model at
finite temperature and nonzero chemical potential with three quark flavors. Chiral and
deconfinement phase transitions are discussed, and the relevant order-like parameters are
analyzed. 
A special attention is  payed  to the critical end point (CEP): 
the influence of the strangeness on the location of the CEP is studied;
also the strength of the flavor-mixing interaction alters the CEP location, once when 
it becomes weaker the CEP moves to low temperatures and can even disappear.

\end{abstract}


Understanding the QCD phase structure is one of the most important topics in the physics of strong
interactions. The developed effort on both the theoretical point of view (by using effective models
and lattice calculations) and the experimental point of view (it is one of the main goals of the heavy
ion collisions program \cite{SQM}) has proved very fruitful, shedding light on properties of matter 
at finite temperatures and densities.

Confinement and chiral symmetry breaking are two of the most important 
features of quantum chromodynamics (QCD) which is the fundamental theory 
of strong interactions. 
Its basic constituents are quarks and gluons that are confined in hadronic matter. 
At high temperatures and densities hadronic matter should undergo a phase transition 
into the quark-gluon plasma (QGP). 
A challenge of theoretical  studies based on QCD is to  predict the
equation of state, the critical end point and the nature of the phase transition.

The fundamental theory of QCD is known to be highly difficult for the analytical analysis 
because of the strong coupling regime. 
Meanwhile, as this regime triggers the most interesting phenomena in QCD, it has become a 
common practice to replace the low-energy QCD by some effective models. 
Examples of such descriptions include NJL  type models which  have been developed providing 
guidance and information relevant to observable experimental signs of deconfinement and QGP features.

NJL type models, that take into account only quark degrees of freedom, give the correct
chiral properties; static gluonic degrees of freedom are then introduced in the NJL
Lagrangian through an effective gluon potential in terms of Polyakov loop with the aim of
including features of both chiral symmetry breaking and deconfinement.

Our calculations are performed in the framework of an extended SU(3)$_f$ PNJL
Lagrangian, which includes the 't Hooft instanton induced interaction term that breaks
the U$_A$(1) symmetry, and the quarks are coupled to the (spatially constant) temporal
background gauge field $\Phi$ \cite{Fukushima,Ratti}:
%
\begin{eqnarray}\label{eq:lag} {\mathcal L}&=& \bar q(i \gamma^\mu D_\mu-\hat m)q +
\frac{1}{2}\,g_S\,\,\sum_{a=0}^8\, [\,{(\,\bar q\,\lambda^a\, q\,)}^2\, +
{(\,\bar q \,i\,\gamma_5\,\lambda^a\, q\,)}^2\,] \nonumber\\
&+& g_D\,\left\{\mbox{det}\,[\bar q\,(1+\gamma_5)\,q] +\mbox{det} \,[\bar q\,(1-\gamma_5)\,q]\right\}
- \mathcal{U}\left(\Phi[A],\bar\Phi[A];T\right).
\end{eqnarray}

The covariant derivative is defined as $D^{\mu}=\partial^\mu-i A^\mu$, with
$A^\mu=\delta^{\mu}_{0}A_0$ (Polyakov gauge); in Euclidean notation $A_0 = -iA_4$.  The
strong coupling constant $g$ is absorbed in the definition of $A^\mu(x) = g {\cal
A}^\mu_a(x)\frac{\lambda_a}{2}$, where ${\cal A}^\mu_a$ is the (SU(3)$_c$) gauge field
and $\lambda_a$ are the (color) Gell-Mann matrices.

The  effective potential for the (complex) field $\Phi$ adopted in our pa\-ra\-me\-tri\-za\-tion of
the PNJL model  reads:

\begin{eqnarray}
    \frac{\mathcal{U}\left(\Phi,\bar\Phi;T\right)}{T^4}
    =-\frac{a\left(T\right)}{2}\bar\Phi \Phi +
    b(T)\mbox{ln}\left[1-6\bar\Phi \Phi  + 4(\bar\Phi^3+ \Phi^3)-3(\bar\Phi
    \Phi)^2\right],
    \label{Ueff}
\end{eqnarray}
where
\begin{equation}
    a\left(T\right)=a_0+a_1\left(\frac{T_0}{T}\right)+a_2\left(\frac{T_0}{T}
  \right)^2\,\mbox{ and }\,\,b(T)=b_3\left(\frac{T_0}{T}\right)^3.
\end{equation}

The parameters of the effective potential $\mathcal{U}$ are given by $a_0=3.51$, $a_1=
-2.47$, $a_2=15.2$ and $b_3=-1.75$. These parameters have been fixed in order to
reproduce the lattice data for the expectation value of the Polyakov loop and QCD
thermodynamics in the pure gauge sector.
When quarks are added, the parameter $T_0$, the critical temperature for the deconfinement 
phase transition (that manifests itself as a breaking of the center symmetry)
within a pure gauge approach, was fixed to $270$ MeV, according to lattice findings.
This choice ensures an almost exact coincidence between chiral crossover and deconfinement at
zero chemical potential, as observed in lattice calculations.

The parameters of the NJL sector are: $m_u = m_d = 5.5$~MeV,
$m_s = 140.7$ MeV, $g_S\Lambda^2 = 3.67$, $g_D \Lambda^5 = -12.36$
and $\Lambda = 602.3$ MeV, which are fixed to
reproduce the values of the coupling constant of the pion, $f_\pi\,=\,92.4$ MeV, and the
masses of the pion, the kaon, the $\eta$ and $\eta^\prime$, respectively,
$M_\pi\,=\,135$ MeV, $M_K\,=\,497.7$ MeV, $M_\eta\,=\,514.8$ MeV and
$M_{\eta^\prime}\,=\,960.8$ MeV.

The inclusion of the Polyakov loop effective potential ${\cal U}(\Phi,\bar\Phi;T)$, that
can be seen as an effective pressure term mimicking the gluonic degrees of freedom of
QCD, is required to get the correct limit. Indeed in the NJL model the ideal gas limit is
far to be reached due to the lack of gluonic degrees of freedom.

The location and even the existence of the CEP in the phase diagram is a matter of
debate. While different lattice calculations predict the existence of a CEP \cite{Fodor},
the absence of the CEP in the phase diagram is still a possibility as was seen in some 
lattice QCD results \cite{deForcrand}, where the first order phase transition region near 
$\mu=0$ shrinks in the quark mass in $\mu$ space when $\mu$ is increased \cite{deForcrand}.

\begin{figure}[t]
\begin{center}
\includegraphics[width=0.65\textwidth]{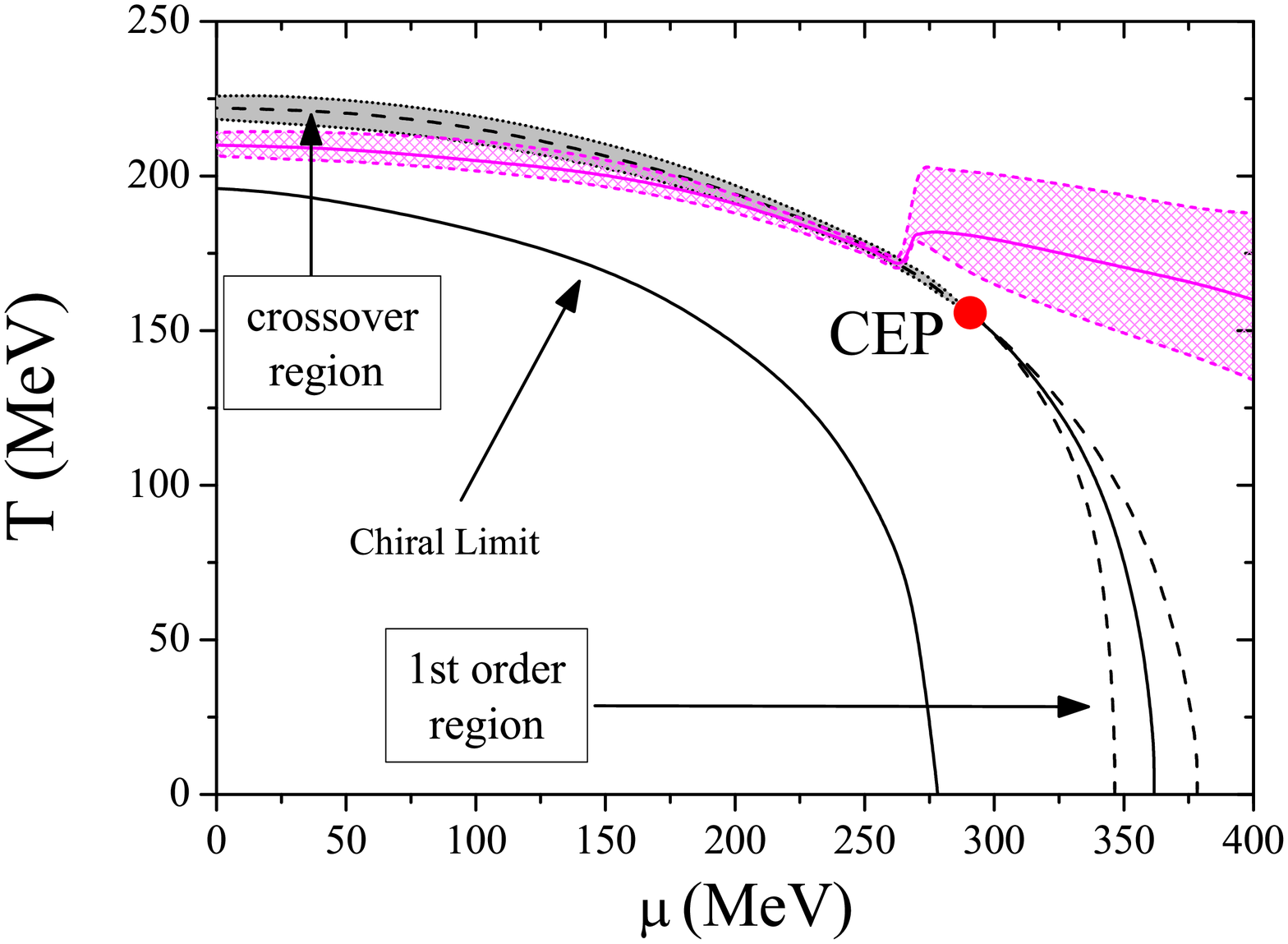}
\end{center}
\caption{The phase diagram in the SU(3) PNJL model: the location 
of the CEP is found at $T^{CEP}=155.80$ MeV and $\mu^{CEP} = 290.67$ MeV 
(see details in text).}
\label{Fig:Phase_Diagram}
\end{figure}

In Figure \ref{Fig:Phase_Diagram} is presented the phase diagram in the PNJL model.
As the temperature increases the chiral transition is a first order one and persists up to the CEP. 
At the CEP the chiral transition becomes a second order one.
The location of the CEP is found at $T^{CEP}=155.80$ MeV and $\mu^{CEP} = 290.67$ MeV
($\rho_B^{CEP}=1.87\rho_0$).
For temperatures above the CEP there is a crossover whose location is calculated making use of  
$\partial^2\left\langle \bar{q}q\right\rangle/\partial T^2=0$,
\textit{i.e.} the inflection point of the quark condensate $\left\langle \bar{q}q\right\rangle$. 

The transition to the deconfinement is given by $\partial^2\Phi/\partial T^2=0$,
and is represented by the magenta line. The surrounding shaded area  that limits
the region where the crossover takes place is determined as the inflection point of the
of the susceptibility $\partial\Phi/\partial T$.   

Due to the importance of the location of the CEP from the experimental point of view,
let us investigate the influence of several parameters which can lead to a significant 
change in the CEP's localization. We study the influence of strangeness on
the location of the CEP (or tricritical point (TCP) in the chiral limit). 

\begin{figure}[t]
\includegraphics[width=0.55\textwidth]{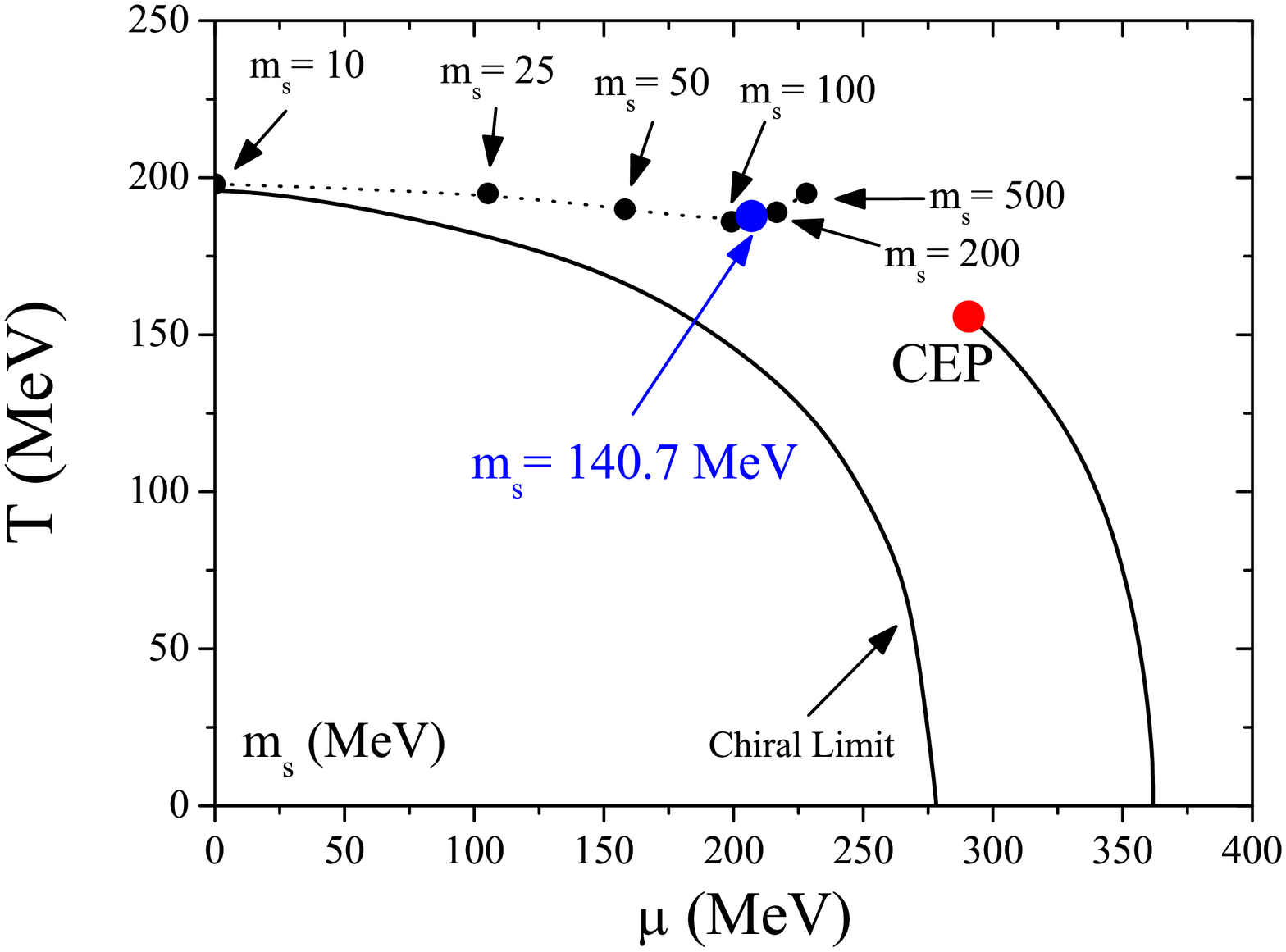}
\hspace{-0.75cm}\includegraphics[width=0.55\textwidth]{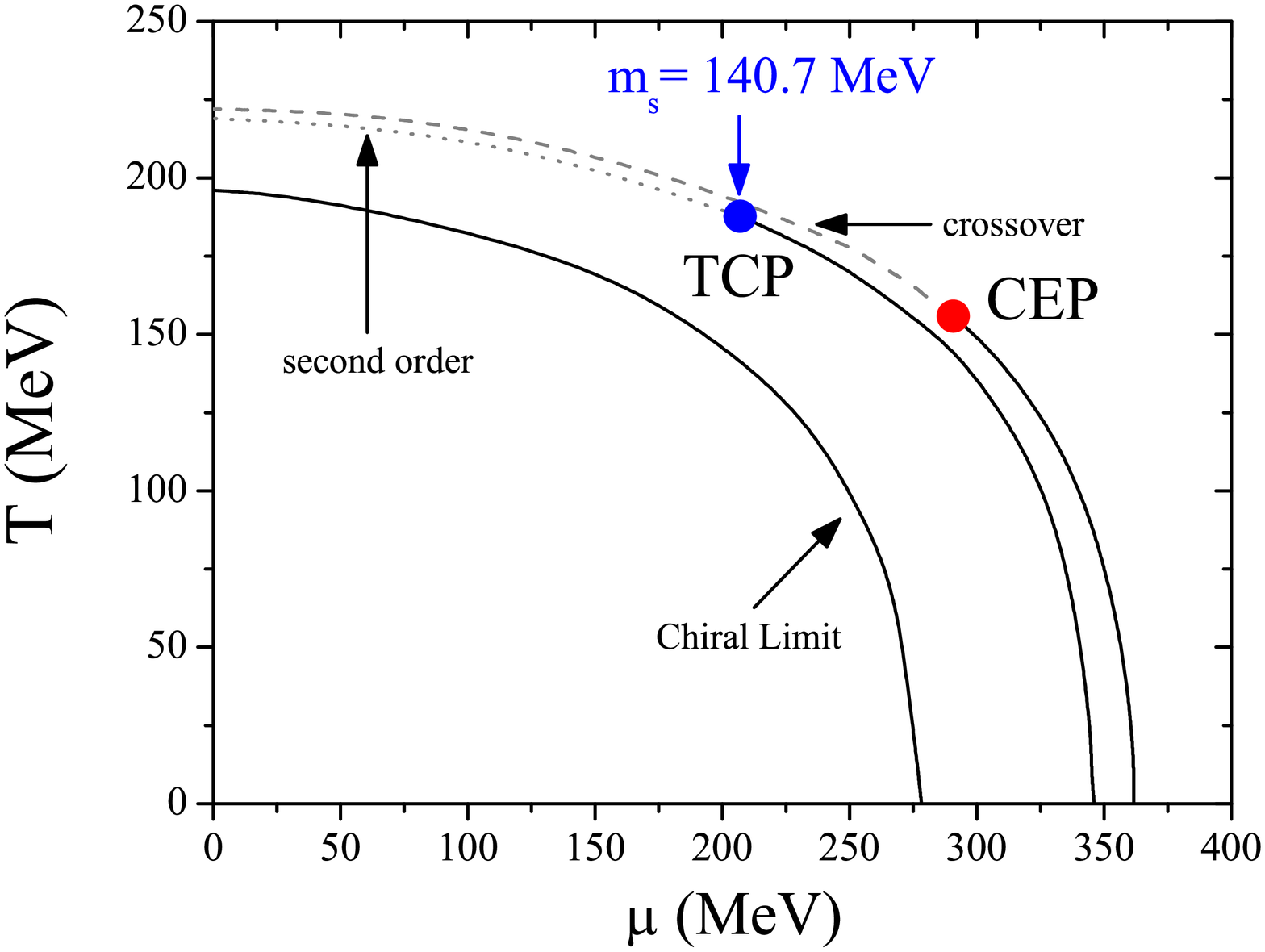}
\caption{Left panel: the phase diagram
in the SU(3) PNJL model. The solid lines represent the first order
phase transition, the dotted line the second order phase transition,
and the dashed line the crossover transition. Right panel: the phase
diagram and the ``line'' of TCPs for  different values of $m_s$ (the
dotted lines are just drawn to guide the eye); the  TCPs in both
figures are obtained  in the limit $m_u=m_d=0$ and $m_s\neq 0$.}
\label{Fig:diagfases}
\end{figure}

A first point to be noticed is that in the PNJL model, when the full chiral limit is considered
($m_u=m_d=m_s=0$), the phase diagram does not exhibit a TCP: chiral
symmetry is restored via a first order transition for all baryonic
chemical potentials and temperatures (see left panel of Figure
\ref{Fig:diagfases}). 
On the contrary, in the chiral limit only for the SU(2) (light) sector
($m_u=m_d=0$, $m_s\neq 0$) a TCP can be found \cite{varios}.
Both situations are in agreement with what is
expected: the chiral phase transition at the chiral limit is of
second order for $N_f = 2$ and first order for $N_f\geq3$
\cite{Pisarski:1983ms}.

To study the influence of strangeness on the location of the
critical points, we vary the current quark mass $m_s$, keeping the
SU(2) sector in the chiral limit and the other model parameters
fixed. In Figure \ref{Fig:diagfases} (left panel) we plot the phase 
diagram of the model in the ($T$, $\mu$) plane,
for various values of the the current quark mass $m_s$.

The pattern of chiral symmetry restoration via first order phase
transition remains for $m_u=m_d=0$ and $m_s<m_{s}^{crit}$. 
The value  for $m_s^{crit}$ is a subject of debate; 
we found $m_s^{crit}\approx 9$ MeV in our model, lower than lattice 
values \cite{Laermann:2003cv} and half of the value obtained in NJL 
model ($m_s^{crit} = 18.3$ MeV \cite{varios}).
When $m_s\geq m_{s}^{crit}$, at $\mu=0$, the transition is a
second order one and, as $\mu$ increases, the line of the second
order phase transition will end in a first order line at the TCP.
Several TCPs are plotted  for different values of $m_s$ in the right
panel of Figure \ref{Fig:diagfases}. As $m_s$ increases, the value
of $T$ for this ``line'' of TCP's decreases as $\mu$ increases
getting closer to the CEP and, when $m_{s}=140.7$ MeV, it starts to
move away from the CEP. The TCP for $m_{s}=140.7$ MeV is the closest
to the CEP and is located at $\mu^{TCP}=206.95$ MeV and
$T^{TCP}=187.83$ MeV. If we choose $m_u=m_d\neq0$, instead of a
second order transition we have a smooth crossover for all the
values of $m_s$ and the ``line'' of TCPs becomes a ``line'' of CEP's.

\begin{figure}[t]
\begin{center}
\includegraphics[width=0.65\textwidth]{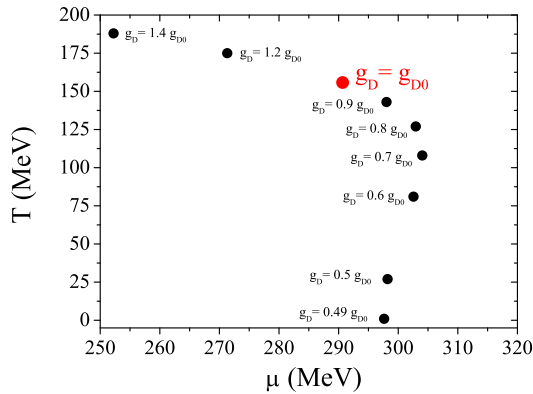}
\end{center} 
\caption{Dependence of the location of the CEP on the strength of the 't Hooft 
coupling constant $g_D$.}
\label{Fig:diagfasesUA1}
\end{figure}

Also the change of the U$_A$(1) anomaly strength has a strong influence on the 
localization of the CEP in the $(T,\,\mu)$ plane. 
In Figure \ref{Fig:diagfasesUA1}, we show the location
of the CEP for several values of $g_D$ compared to the results for $g_{D_0}$, 
the value used for the vacuum. 
As already pointed out by K. Fukushima in \cite{Fukushima}, we also
observe that the location of the CEP depends on the value of $g_D$. 
In fact,  our results show that the existence or not of the CEP is determined 
by the strength of the anomaly coupling, the CEP getting closer to the 
$\mu$ axis as $g_D$ decreases.

As a conclusion, we investigated the phase diagram of the so-called PNJL model 
at finite temperature and nonzero chemical potential with three quark flavors. 
Chiral and deconfinement phase transitions are discussed, and the relevant 
order-like parameters are analyzed.

The chiral phase transition is a first order one in the chiral limit.
Working out of the chiral limit, at which both chiral and center symmetries 
are explicitly broken, a CEP which separates first and crossover lines is found,
and the corresponding order parameters are analyzed.

A special attention is  payed  to the critical end point.
We studied the influence of strangeness on the location of the
critical points varying the current quark mass $m_s$, keeping the
SU(2) sector in the chiral limit and the other model parameters
fixed. 
When $m_s\geq m_{s}^{crit}$, at $\mu=0$, the transition is of the
second order and, as $\mu$ increases, the line of the second
order phase transition will end in a first order line at the TCP.
If we choose $m_u=m_d\neq0$, instead of a
second order transition we have a smooth crossover for all the
values of $m_s$ and the TCP becomes a CEP which depend strongly on the
the value of $m_s$.

We also analyzed the effect of the anomalous coupling strength on the location of the CEP. 
We showed that the location of the CEP depends on the value of $g_D$ and, 
as this strength of the flavor-mixing interaction becomes weaker, the CEP moves 
to low temperatures and can even disappear.

\vskip0.25cm
Work supported by  Centro de F\'{\i}sica Computacional and F.C.T. under Project
No. CERN/FP/116356/2010.


\end{document}